\documentclass[%
 reprint,
superscriptaddress,
 amsmath,amssymb,
 aps,
]{revtex4-1}

\usepackage[pdftex]{graphicx, color}
\usepackage{multirow,booktabs}
\usepackage{bm}
\usepackage{braket}
\usepackage{siunitx}
\usepackage{physics}
\usepackage{hyperref}
\usepackage{ulem}
\usepackage{comment}
\usepackage{here}
\usepackage{mathrsfs}

\newcommand{\xx}{ {\bm x }}
\newcommand{\uu}{ {\bm u }}

\newcommand{\rr}{{\bm r}}
\newcommand{\kk}{{\bm k}}

\newcommand{\pp}{ {\bm p }}

\newcommand{\beb}{\begin{itembox}}
\newcommand{\enb}{\end{itembox}}

\newcommand{\e}{\varepsilon}
\newcommand{\E}{{\bm E}}

\newcommand{\average}[1]{\ensuremath{\left< #1 \right> }}
\newcommand{\1}{\mbox{1}\hspace{-0.25em}\mbox{l}}

\begin{document}

\title{Anomalous Hydrodynamic Transport in Interacting Noncentrosymmetric Metals}

\author{Riki Toshio}
\email[]{toshio.riki.63c@st.kyoto-u.ac.jp}
\affiliation{%
 Department of Physics, Kyoto University, Kyoto 606-8502, Japan
}%

\author{Kazuaki Takasan}
\affiliation{%
  Department of Physics, University of California, Berkeley, California 94720, USA
}%

\author{Norio Kawakami}
\affiliation{%
 Department of Physics, Kyoto University, Kyoto 606-8502, Japan
}%


\date{\today}

\begin{abstract}
In highly conductive metals with sufficiently strong momentum-conserving scattering, the electron momentum is regarded as a long-lived quantity, whose dynamics is described by an emergent hydrodynamic theory. 
In this paper, we develop an electron hydrodynamic theory for noncentrosymmetric metals, where a novel class of electron fluids is realized by lowering crystal symmetries and the resulting geometrical effects. The obtained hydrodynamic equation suggests a nontrivial analogy between electron fluids in noncentrosymmetric metals and chiral fluids in vacuum, and predicts novel hydrodynamic transport phenomena, that is, \textit{asymmetric Poiseuille flow} and  \textit{anomalous edge current}. Our theory also gives a hydrodynamic description of the counterpart of various anomalous transport phenomena such as the quantum nonlinear Hall effect. Furthermore, we give a symmetry consideration on the hydrodynamic equation and propose several experimental setups to realize such anomalous hydrodynamic transport in the existing hydrodynamic materials, including bilayer graphene and Weyl semimetals.

\end{abstract}

\pacs{Valid PACS appear here}

\maketitle



\textit{Introduction.}---
How to understand electron dynamics in crystals has been a fascinating problem in condensed matter physics. In particular, geometrical and topological contributions to the dynamics have attracted much interest in recent decades and various anomalous transport phenomena have been investigated~\cite{Xiao2010, Nagaosa2010}. However, while our understanding of the dynamics has become highly sophisticated at the level of single-particle approximation, we still do not understood well, in strongly correlated systems, what kind of role the electron correlation plays in the transport and what paradigm enables us to describe the complex nonlinear and nonlocal dynamics, including the geometrical aspects, beyond the single-particle picture. 

Nowadays, the hydrodynamic theory is considered as a promising framework to analyze such a dynamics~\cite{Zaanen2016, Lucas2018_review, Polini2019, Andrea2019}. It is a general framework for describing the low-energy dynamics in interacting many-particle systems~\cite{Landau_Fluid,Chaikin} and believed applicable to highly conductive metals where a large separation of scales between momentum-relaxing and momentum-conserving scattering is realized (we refer to such materials as ``hydrodynamic materials''). 
In fact, recently, many pieces of evidence for hydrodynamic electron flow have been reported through unconventional DC transport phenomena, in various materials such as GaAs quantum wells~\cite{Molenkamp1994,Molenkamp1995, Braem2018, Gusev2018-2, Gusev2018, Levin2018}, 2D monovalent layered metal PdCoO${}_2$~\cite{Moll2016}, Weyl semimetal WP${}_2$~\cite{Gooth2018}, and monolayer (ML)/bilayer (BL) graphene~\cite{Bandurin2016, Crossno2016, Kumar2017, Bandurin2018,  Sulpizio2019, Berdyugin2019}. Furthermore, it has been pointed out that the hydrodynamic regime is also realized in a noncentrosymmetic metal MoP~\cite{Kumar2019}.

One of the remarkable differences between the usual fluids, such as water, and the electron fluids in crystals is that the latter is living in the background of the crystal lattice and thus always reflects the crystal symmetries. Especially in the noncentrosymmetric metals, a finite Berry curvature in $\kk$-space emerges due to the symmetry reductions, and therefore noncentrosymmetric hydrodynamic materials, such as BL graphene, WP${}_2$ and MoP, are expected to show anomalous hydrodynamic transport phenomena. This fact is also closely related to the multi-band nature of the electrons, which highlights a crucial difference between electron fluids in crystals and those in vacuum. 
However, despite the recent intensive research activities, symmetry or geometry considerations on the electron fluids have not been executed so far, except for several limited problems~\cite{Gorbar2018, Lucas2019, Rao2019, Georgious2020}.

In this Letter, we formulate the hydrodynamic theory for time-reversal-symmetry~(TRS) preserved noncentrosymmetric metals, which is correct up to the second order in electric fields. The obtained equations reveal the emergence of additional anomalous forces in the Euler equation, which suggests a nontrivial analogy with the inviscid chiral fluids in quark-gluon plasma. Especially remarkable is that the analogy indicates the existence of {\it vorticity-induced electric current} in condensed matter without chiral anomaly, which is analogous to the chiral vortical effect in the chiral fluids. 
We predict that this anomalous effect, together with the viscosity effect, causes a novel anomalous hydrodynamic flow, that is, \textit{asymmetric Poiseuille flow} and \textit{anomalous edge current}. Furthermore, our theory gives a clear description of the hydrodynamic counterpart of a variety of anomalous nonlinear electric and thermoelectric transport phenomena, such as the quantum nonlinear Hall effect (QNHE)~\cite{Sodemann2015} and the magnus Hall effect~\cite{Papaj2019}. These responses have been formulated well in the ohmic or the ballistic regime, but not in the hydrodynamic regime so far. 
Finally, we give a symmetry consideration for the existing hydrodynamic materials and estimate the order of the anomalous contributions using an effective model for TMDs.

\textit{Derivation of hydrodynamic equations.}---
We outline how to derive the hydrodynamic equations for the TRS-preserved noncentrosymmetic metals (For detail, see Supplemental Materials). We start from the Boltzmann equation and the semi-classical equations~\cite{Xiao2010} 
\begin{equation}
	\hbar{\dot \kk}_\alpha =-e \E,\ \ \ 
	{\dot \rr}_\alpha=\frac1\hbar \pdv{{\e}_\alpha(\kk)}{\kk} -{\dot \kk}_\alpha\times
	{\bm \Omega}_\alpha(\kk),
\end{equation}
where $\alpha$ and $\e_\alpha$ are a band index and the energy dispersion. ${\bm \Omega}_\alpha$ is the Berry curvature of the electrons, defined as ${\bm \Omega}_\alpha\equiv\grad_\kk\times {\bm A}_\alpha$ with ${\bm A}_\alpha\equiv
i\average{u_{\alpha\kk}|\grad_\kk u_{\alpha\kk}}$.
In particular, it should be noted that the Berry curvature appears only when the systems break the time-reversal or inversion symmetry. In the following, we restrict the discussion to the materials whose band structure near the Fermi level is composed of several equivalent valleys having an isotropic parabolic dispersion with mass $m$. Such a condition is realized in the materials such as graphene with inversion breaking and monolayer transition metal dichalcogenides (ML-TMDs).

Following the standard approach~\cite{Landau_Kinetics, Lucas2018_review}, the continuity equation for momentum is obtained by multiplying the Boltzmann equation with momentum, and integrating the momentum in the relaxation time approximation for the momentum-relaxing scatterings. Especially in the hydrodynamic regime, we can express the momentum flux explicitly in terms of the hydrodynamic variables, temperature $T$, chemical potential $\mu$ and velocity $\uu$, and reach the generalized Euler equation for noncentrosymmetric metals up to the  second-order perturbation in $\uu$ and $\E$ as follows:
\begin{equation}\label{Euler}
\begin{aligned}  
	&\pdv{\uu}{t}+({\bm u}\cdot \grad)\uu +\frac{\grad P}{\rho}  + \frac{e}{n\hbar} \left[\frac1{m}\hat{C}(\curl \E) \right. \\
  & \left. + \hat{F} \left(\E\times \frac{\grad T}{T}\right) +\hat{D}(\E\times \grad \mu)\right]
	+ \frac{e}{m}\E  =-\frac{\uu}{\tau_{mr}},
\end{aligned}
\end{equation}
where $n$ and $\rho$ are the density of particles and mass, $P$  is the pressure, $\E$ is an applied electric field, $\tau_{mr}$ is the momentum relaxation time. 
Here, $\hat{D}$ is a second-rank pseudotensor, so-called {\it Berry curvature dipole}~\cite{Sodemann2015} 
\begin{equation}
D_{il} = \sum_\alpha D_{il}^\alpha ,\ \  D_{il}^\alpha \equiv -\int [d\pp] \Omega_{\alpha,l} \pdv{f_{0\alpha}}{p_i},
\end{equation}
and, $\hat{C}=\sum_\alpha \hat{C}^\alpha$ and $\hat{F}=\sum_\alpha \hat{F}^\alpha$ are different geometrical coefficients, defined as
\begin{equation}
	C_{il}^\alpha \equiv \int [d\pp]\ p_i  \Omega_{\alpha,l} f_{0\alpha},\ 
F_{il}^\alpha \equiv -\int [d\pp]  \e_\alpha \Omega_{\alpha,l} \pdv{f_{0\alpha}}{p_i},
\end{equation}
where $f_{0\alpha}= [1+e^{-\beta(\e_\alpha(\pp)-\mu )}]^{-1}$ is the Fermi distribution function for the valley $\alpha$ and we introduce the notation $\int [d\pp]\equiv\int d\pp/(2\pi\hbar)^d$. We note that $\pp$ is defined as a deviation from each valley center. This result indicates that the symmetry lowering of the crystal leads to the emergence of novel anomalous forces in the conventional Euler equation and, as seen below,  these forces drive an unusual electric flow accompanied by non-uniformity of hydrodynamic variables. Especially  remarkable is that, even though our fluids have no chiral anomaly, the obtained equation has a quite similar form to that of inviscid chiral fluids, which is believed to realize in quark-gluon plasmas~\cite{Son2009, Hidaka2018}. This fact is expected to pave the way for the realization of the anomalous nonlinear transport analogous to that in chiral fluids, even in condensed matter systems without chiral anomaly.


To relate the hydrodynamic theory with an observable current, we further need to describe the so-called ``transport current''~\cite{Xiao2006} in terms of velocity field $\uu$. Generally, the current is known to be described with orbital magnetization ${\bm M}$ as follows~\cite{Cooper1997, Xiao2006}:
\begin{equation}
{\bm J}=\sum_\alpha \left[\int[d\kk] \dot{\rr} f_\alpha + \curl \int[d\kk]{\bm m}_\alpha f_\alpha \right] -\curl {\bm M}.
\end{equation}
Under a phenomenological assumption on the orbital magnetization~\cite{comment}, we obtain the following hydrodynamic expression for the current in the second order perturbation in $\uu$ and $\E$:
\begin{equation}\label{hydro_current}
\begin{aligned}
			{\bm J} = &-en{\bm u}-\frac e\hbar \left[m (e\E+ \grad \mu) \times ({}^t\!\hat{D} {\bm u})\right.\\
		&\left. +\curl ({}^t\!\hat{C} {\bm u})+ m ({\grad T}/{T}) \times ({}^t\!\hat{F} {\bm u})\right],
\end{aligned}
\end{equation}
where the first term is the conventional part appearing in the familiar hydrodynamics, and the others are anomalous parts reflecting the symmetry lowering of the fluids. Here, we note that $\grad$ does not act on $\hat{C}$. Remarkably, the second term in the bracket denotes a generalization of {\it chiral vortical effect} in chiral fluids~\cite{Son2009}, because the term includes an anomalous current induced by the vorticity fields, which supports the above analogy. This phenomenon can be intuitively understood as the correction due to a magnetization current induced by the inhomogeneous velocity fields through the orbital Edelstein effect~\cite{Yoda2015, Yoda2018}. In the following, we refer to this effect as {\it generalized vortical effect} (GVE).


\textit{Results for anomalous transport.}---
Here, we give several demonstrations of the above theory, including optical or thermal responses and finite size effects in the fluids. From the symmetry viewpoint, it is noteworthy that linear and local anomalous responses, such as the anomalous (thermal) Hall effect, are prohibited by TRS, so that nonlocal or nonlinear anomalous responses play a key role in our fluids. 
First, for the simplest problem, let us consider the first- and second- responses to a uniform electric field. In this case, we can express electric fields as $\E=\Re [\tilde{\E}e^{i\omega t}]$ with $\tilde{\E}\in \mathbb{C}$, and the electric current is denoted in the form $J_i = \Re [J_i^0 +  J_i^{\omega}e^{i\omega t}+ J_i^{2\omega} e^{2i\omega t}]$. 
Substituting the solution of Eq.~(\ref{Euler}) to Eq.~(\ref{hydro_current}), we easily reach the following expression of the linear and nonlinear conductivity tensor: 
\begin{equation}
\sigma^{(1)} = \frac{\sigma_D}{1+i\omega\tau_{mr}},\ 
\sigma_{ijk}^{(2)} = -\e_{ilk} \frac{e^3\tau_{mr}}{2(1+i\omega\tau_{mr})}D_{jl},
\end{equation}
where $J_i^\omega =\sigma^{(1)} \tilde{E}_i$, $J_i^0 = \sigma_{ijk}^{(2)} \tilde{E}_j \tilde{E}_k^*$, $J_i^{2\omega} = \sigma_{ijk}^{(2)} \tilde{E}_j \tilde{E}_k$ and $\sigma_D=ne^2\tau_{mr}/m$. The linear one is so-called Drude conductivity, and the nonlinear one agrees with the results of the quantum nonlinear Hall effect (QNHE)~\cite{Sodemann2015}. Under TRS, the QNHE gives a leading anomalous current, which has  already been observed in several materials including hydrodynamic materials such as GaAs quantum well~\cite{Ganichev2001, Olbrich2009, Moore2010} and BL-graphene~\cite{Ho2019}. This result indicates that our theory can correctly reproduce the known results for the uniform electric responses formulated so far in the momentum-dissipative or ballistic regimes, but not in the hydrodynamic regime.

Furthermore, our hydrodynamic theory elucidates the anomalous nonlinear/nonlocal responses under non-uniform hydrodynamic variables. Let us start with the nonlinear thermoelectric transport induced by the coupling between $\grad T$ and $\E$. 
Imposing a uniform thermal gradient and electric field on the fluids, we obtain the following anomalous contribution  proportional to the product of $\grad T$ and $\E$: 
\begin{equation}
{\bm J}' = \frac{e^2}{\hbar T} \frac{\tau_{mr}}{1+i\omega\tau_{mr}}
\left[ 
\hat{F} (\tilde{\E}\times \grad T ) + {\grad T} \times ({}^t\!\hat{F} \tilde{\E})
\right].
\end{equation}
Under TRS, this contribution gives a leading anomalous current under the thermal gradient. Interestingly, the second term is always perpendicular to $\grad T$, whereas the other is directed only by the tensor $\hat{F}$, whose form is determined by the crystal symmetry as shown below. These features enable us to control or switch the current direction by changing either direction of $\grad T$ or $\E$ without changing the other. Similar responses are also discussed in dissipative (ohmic) regime in Ref.~\cite{Nakai2019}.
Similarly, from our theory, we can calculate an anomalous current under chemical potential bias, which is proportional to the product of $\E$ and $\grad \mu$ and described by the Berry curvature dipole as in the case of QNHE. 
This phenomenon is in contrast to the \textit{Magnus Hall effect}, which has been formulated in the ballistic regime \cite{Papaj2019} and describes the Hall current responding to $\E\parallel\grad \mu$. We note that our result is qualitatively different from that in Ref.~\cite{Papaj2019}, since the former describes the transport in the hydrodynamic regime and  includes an anomalous response to $\E\perp\grad \mu$ as well as the correction due to the magnetization current.

\begin{figure}[t]
\centering
\includegraphics[width=5.5cm]{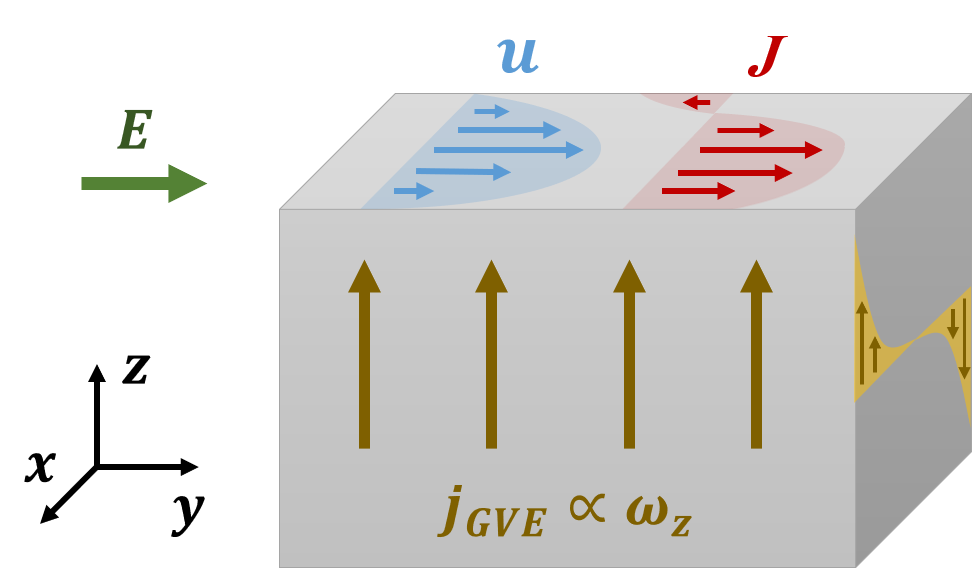}
\caption{Schematic picture of the asymmetric Poiseuille flow (red) and anomalous edge current (yellow) induced by the GVE in a 3D sample with finite width $w$ in the $x$-direction.}
\label{Poi}
\end{figure}

\begin{figure}[t]
  \centering
  \begin{tabular}{ll}
(a)  &(b) \\
   \includegraphics[width=4.2cm]{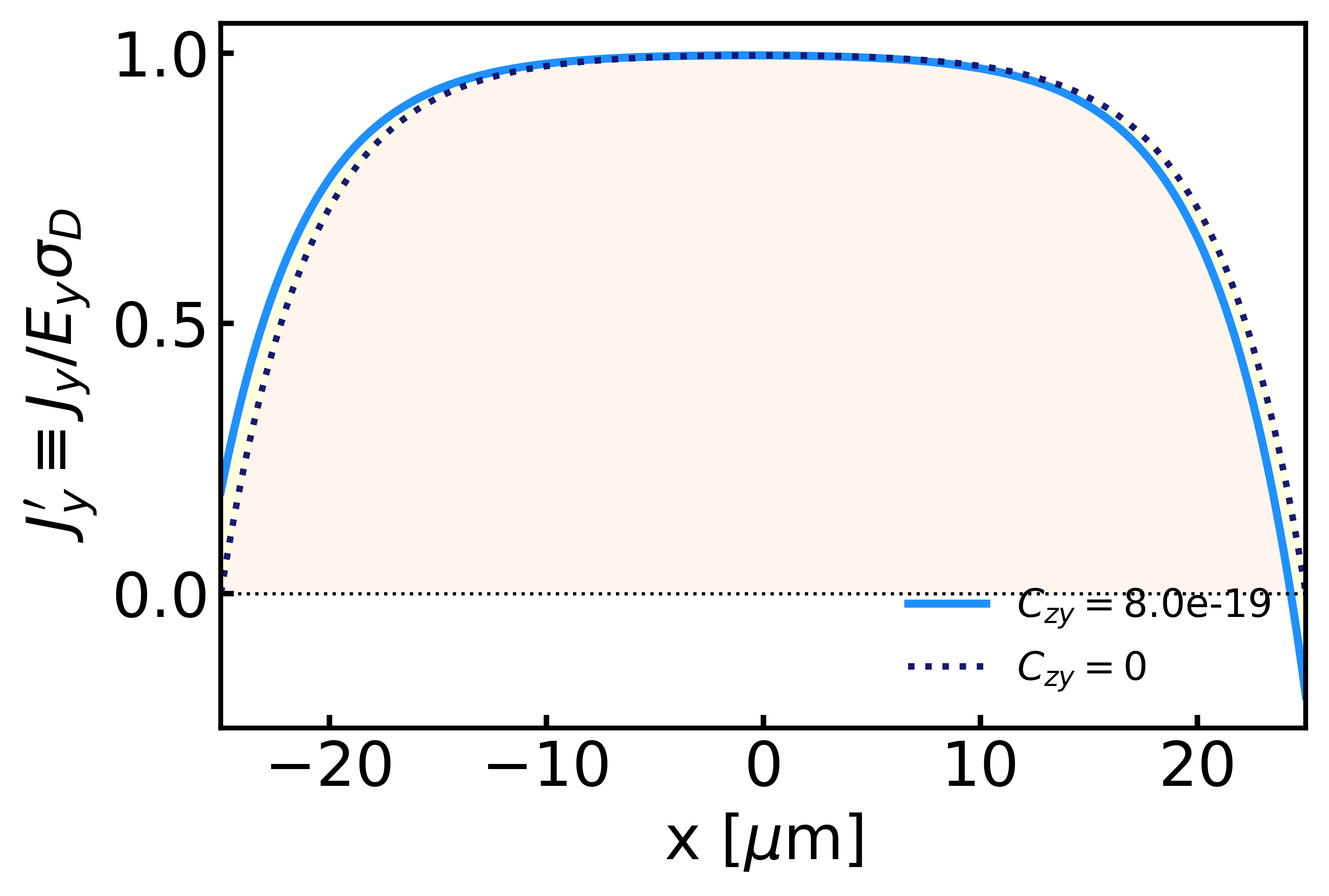}&
    \includegraphics[width=4.2cm]{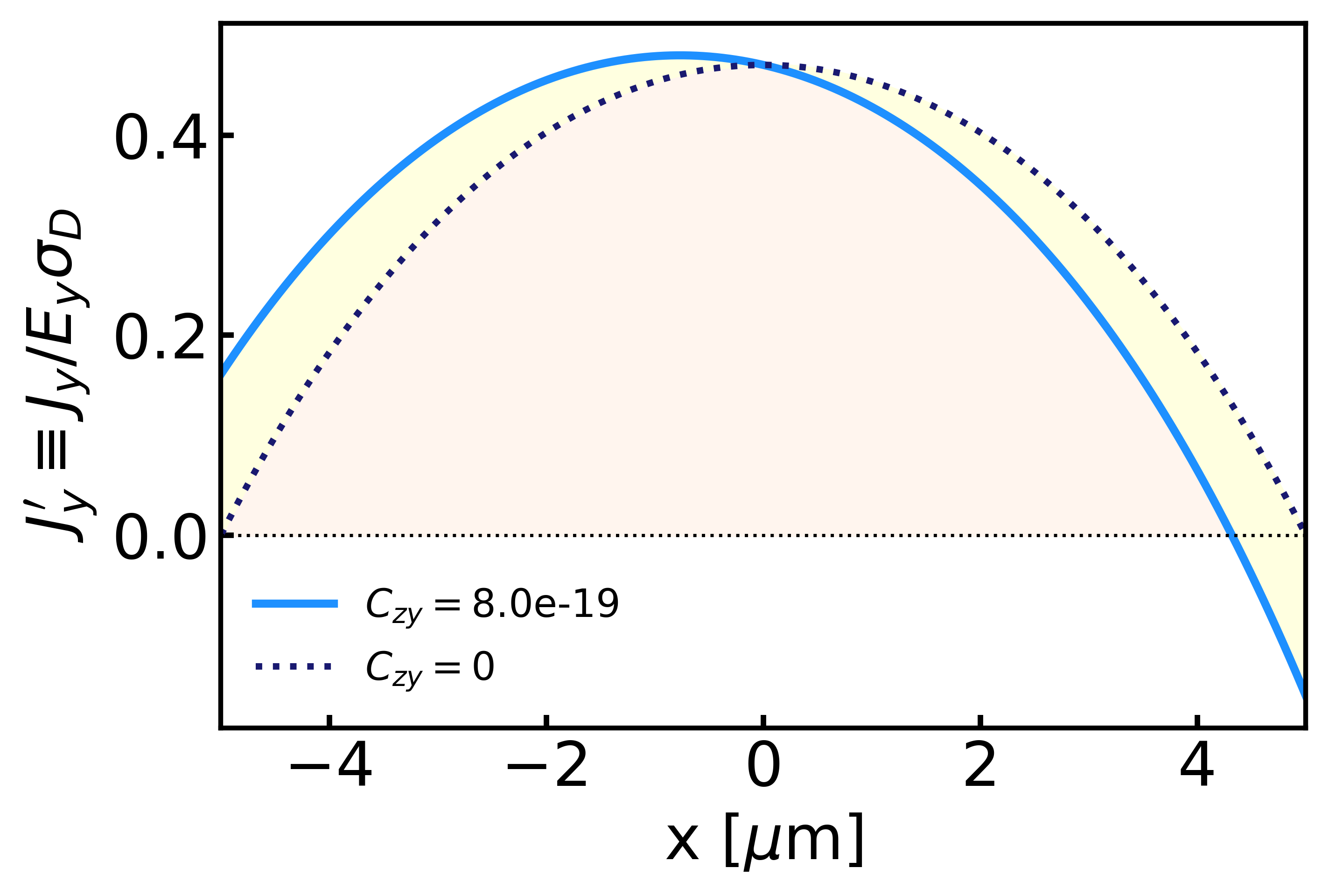}\\
(c) & (d) \\
   \includegraphics[width=4.2cm]{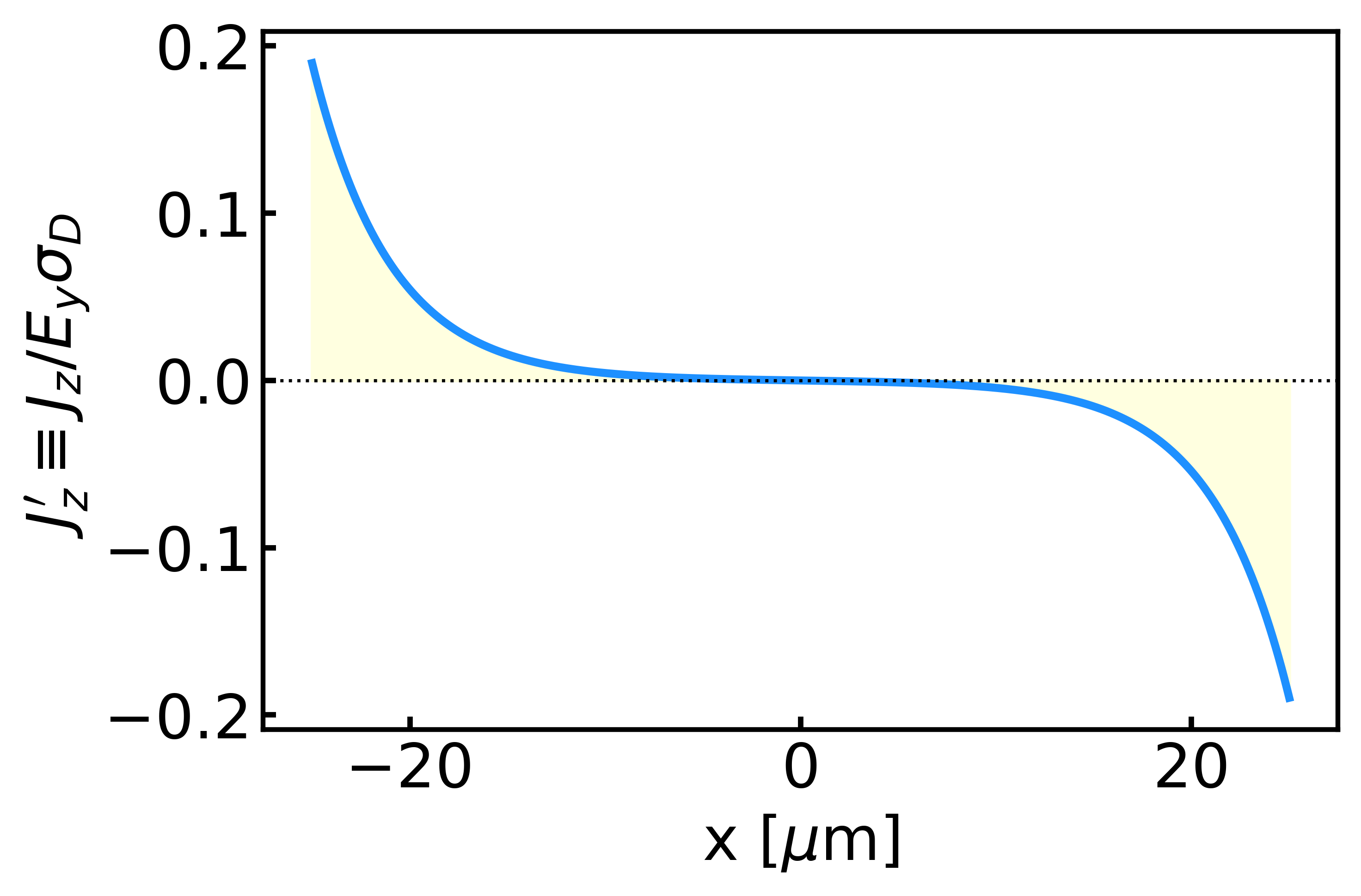}&
    \includegraphics[width=4.2cm]{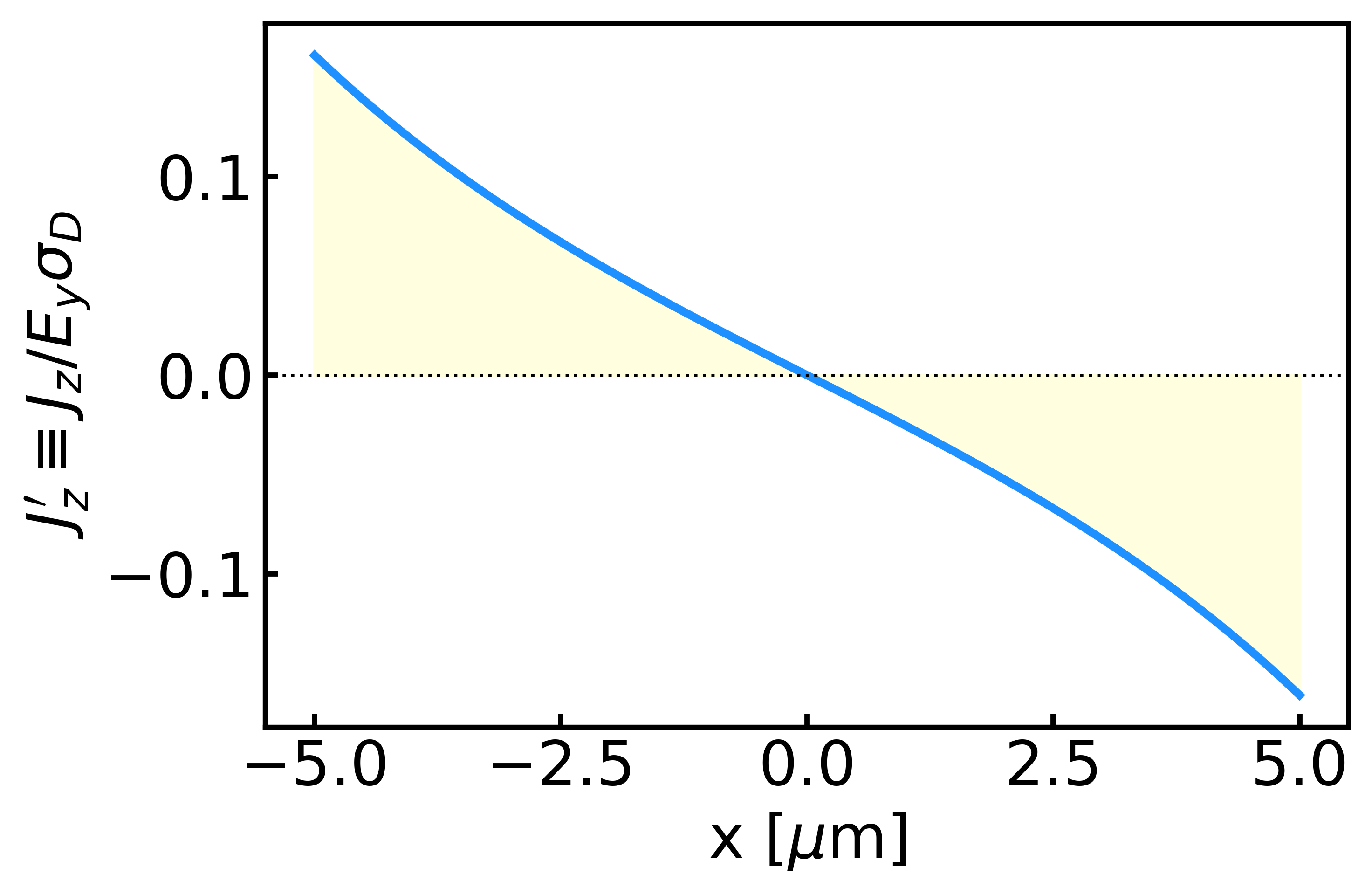}
  \end{tabular}
\caption{Plot of the normalized current profiles $j_y'(x)$ and $j_z'(x)$ in the $x$-range $[-w/2 , w/2]$. (a), (c) : $w=\SI{50}{\micro m}$ and (b), (d) : $w=$\SI{10}{\micro m}. We use the parameters; $\nu=3.8\times 10^{-2}$~m${}^2$/s, $\tau_{mr}=5\times 10^{-10}$~s, $m=1.2m_e$, $C_{yy}=C_{zy}=8.0\times 10^{-19}$~kg/s, where $m_e$ is the bare electron mass. Here, the dotted lines in (a) and (b) show the usual Poiseuille flow realized in centrosymmetric metals ($C_{zy}=0$). }
\label{plot}
\end{figure}

The most significant consequence of our theory is that the combination of the GVE and the viscosity effect gives rise to an unprecedented current flow in finite size systems, which is unexplored so far.  
For the demonstration, we consider the Poiseuille flow in 3D samples with finite width $w$ in the $x$-direction~(Fig.~\ref{Poi}), which  most clearly characterizes the hydrodynamic transport in noncentrosymmetric metals.
Here, to estimate boundary effects, we introduce the viscosity term $\nu \Delta u$ into the Euler equation~(\ref{Euler})~\cite{viscosity}. When we apply an electric field in the $y$-direction, the electron fluids form the Poiseuille flow as in Fig.~\ref{Poi} and the velocity profile is given by
\begin{equation}
u_y(x)=\frac{e\tau_{mr}E}{m}\left[ 1-\frac{\cosh(x/l)}{\cosh(w/2l)}    \right],
\end{equation}
and the vorticity is also calculated as 
\begin{equation}\label{vorticity}
\omega_z(x) = \pdv{u_y}{x} = \frac{e\tau_{mr}E}{ml}  \frac{\sinh(x/l)}{\cosh(w/2l)} ,
\end{equation}
where $-w/2\leq x \leq w/2$ and $l \equiv \sqrt{\nu \tau_{mr} }$. The expression~(\ref{vorticity}) indicates that the vorticity distributes over a width of $l$ from the boundaries. Using Eq.~(\ref{hydro_current}), the electric current is obtained as  
\begin{equation}\label{edge_current}
J_y = -enu_y +(e/\hbar )C_{yz}\omega_z, \  J_z= -(e/\hbar )C_{yy} \omega_z.
\end{equation}
These results demonstrate the realization of two unprecedented phenomena : {\it anomalous edge currents} and {\it asymmetric Poiseuille flow}. Firstly, the second equation indicates that, through the diagonal component $C_{yy}$, the vorticity fields induce the $z$-directed anomalous current localized over a width of $l$ from the boundaries~(Fig.~\ref{Poi}), which is directed oppositely at two sides~\cite{Yip}. 
Secondly, the first equation means that the off-diagonal component $C_{yz}$ causes an antisymmetric current distribution in the $x$-direction, which leads to the realization of an asymmetric $x$-distribution of the total current, i.e. {\it asymmetric Poiseuille flow}. 
In particular, it is noteworthy that, in the anomalous flow, there arises a backflow of electric current at one side, which is directed against the electric field. 
For example, as we discuss below, these phenomena could be realized in the hydrodynamic material such as WP${}_2$, where we can estimate the length $l$ as $w=\SI{4}{\micro m}$, using the typical values $\nu=3.8\times 10^{-2}$~m${}^2$/s, $\tau_{mr}=5\times 10^{-10}$~s in the experiments at 4K~\cite{Gooth2018}. In Fig.~\ref{plot}, we plot the current profile for width $w=\SI{50}{\micro m}$ and $w=\SI{10}{\micro m}$ with the above values of the parameters. 
Another possibility to detect the GVE is the measurement of the nonlocal resistivity in the so-called vicinity geometry~\cite{Iacopo2015, Bandurin2016}. In this case, a large velocity gradient (especially, whirlpool) is formed around a narrow current injector, and thereby, the GVE is expected to give a finite contribution to the nonlocal resistivity.

\textit{Symmetry consideration.}---
Here, we provide a symmetry consideration on our hydrodynamic theory for two-dimensional (2D) and three-dimensional (3D) systems. The coefficients $\hat{D}$, $\hat{C}$, and $\hat{F}$ are second rank pseudo-tensors with the same symmetry, and known to become finite in the materials characterized by the so-called natural optical activity~\cite{Landau_Electromagnetics, Halasyamami1998, Puggioni2014}.
Based on the group theory, in the following, we give the symmetry classification and 
discuss how we can realize anomalous transport in the existing hydrodynamic materials. The results are summarized in Table~\ref{table}.

(i) For 3D systems: As for the antisymmetric part of $\hat{A}$ (referred to as $\hat{A}^-$), we can represent it in terms of the dual polar vector $a_i\equiv \epsilon_{ijk}A^-_{jk}/2$. This vector is allowed to be finite only in the polar point group $\{C_n, C_{nv}\}$, with $n=1,2,3,4,6$, and oriented along the polar axis. With this vector representation, for example, we can rewrite the second term in the bracket of Eq.~(\ref{Euler}) as 
\begin{equation}
\hat{F}^-\left(\E\times {\grad T}/{T}\right)= -{\bm f}\times \left(\E\times {\grad T}/{T}\right),
\end{equation}
and the transport current~(\ref{hydro_current}) as
\begin{equation}
\begin{split}
		{\bm J}^- &=  -\frac e\hbar  \left[
		m(e\E+\grad\mu)\times \left({\bm d} \times{\bm u} \right) + ({\bm c}(\div\uu) \right. \\
		&\left.-({\bm c}\cdot \grad)\uu )+ m({\grad T}/{T}) \times \left( {\bm f}\times \uu \right)
		\right].
\end{split}
\end{equation}
In particular, the asymmetric Poiseuille flow is realized through this asymmetric contribution.  
As for the existing hydrodynamic materials, WP${}_2$, which belongs to the polar point group $C_{2v}$~\cite{Kumar2017MR}, can take a finite value of ${\bm a}$. On the other hand, the other 3D hydrodynamic materials PtSn${}_4$ and MoP are prohibited to have finite ${\bm a}$, each of which belongs to the point group $D_{2h}$ and $D_{3h}$~\cite{Fu2018,Kumar2019}. 

Meanwhile, the symmetric part of $\hat{A}$ is allowed in any chiral groups $\{ O, T,C_1,C_n,D_n\}$, with $n=2,3,4,6$, and specific non-chiral groups $\{C_{s}, C_{2v}, D_{2d}, S_4\}$, and it induces the anomalous edge current~(\ref{edge_current}) in the Poiseuille flow. 
Especially the scalar component $C\equiv \mathrm{Tr}[\hat{C}]$ contributes to the GVE term in Eq.~(\ref{hydro_current}) in the form $C{\bm \omega}$ $({\bm \omega} \equiv \curl \uu)$, which causes phenomena similar to the CVE in chiral fluids. However, although such a trace component is allowed to be finite in the chiral point group from a symmetry perspective, we can easily show that $C$ is always zero under our approximation of parabolic dispersion~\cite{comment}. 

\begin{table}[b] \label{table}
\caption{Symmetry constraints on the candidates for hydrodynamic materials. For comparison, we also refer to some TMD materials. Here, ${\bm m}$ and $\hat{A}$ represent the vector perpendicular to a mirror plane and either of $\hat{D}$, $\hat{C}$, and $\hat{F}$. The line ``operations'' denotes the operations needed to make $\hat{A}$ finite. For  the symmetry consideration on BL-graphene and ML-MoS${}_2$, implicitly, we also take into account the polar effect of the substrate: $D_{3h} \to C_{3v}$.}
\begin{center}\label{table}
\begin{tabular}{|l|c|c|c|c}
\hline
Material & point group & components  & operations \\
\hline
MoP & $ D_{3h}$ & NO &  ---\\
WP${}_2$ & $C_{2v}$ & $A_{xy}, A_{yx}$ &---  \\
PdCoO${}_2$ & $D_{3d}$ & NO &  ---\\
\hline
(110)-GaAs  & $C_{s}$  & ${\bm A} \parallel {\bm m}$ &--- \\
ML-Graphene &$D_{6h}$ & NO & ---\\
BL-Graphene & $D_{3h}\to C_{s}$ & ${\bm A} \parallel {\bm m}$ & uniaxial strain \\
ML-WTe${}_2$ & $C_{s}$ & ${\bm A} \parallel {\bm m}$ &---  \\
ML-MoS${}_2$ & $D_{3h}\to C_{s}$ & ${\bm A} \parallel {\bm m}$ & uniaxial strain \\

\hline
\end{tabular}
\end{center}
\label{default}
\end{table}%

(ii) For 2D systems: Next we consider 2D materials including the existing hydrodynamic materials such as mono/bi-layer graphene and GaAs quantum well. In this case, since the Berry curvature behaves as a pseudo-scalar, the coefficients $\hat{A}$ behave as pseudo-vectors constrained in the 2D plane: $A_{ij}=A_i\delta_{jz}$. 
The symmetry constraints force the vector ${\bm A}$ to be orthogonal to the mirror lines, and thus, 
${\bm A}$ is allowed only in the crystals with less than two mirror lines in the plane. 
With this vector representation, we can rewrite the second terms in the bracket of Eq.~(\ref{Euler}) as 
\begin{equation}
\hat{F}\left(\E\times {\grad T}/{T}\right)=  \left(\E\times {\grad T}/{T}\right)_z{\bm F} ,
\end{equation}
and the transport current~(\ref{hydro_current}) as
\begin{equation}
\begin{aligned}
			{\bm J}& = -en{\bm u}-\frac e\hbar \left[m ({\bm D} \cdot \uu) \cdot(e\E+ \grad \mu) \times \hat{{\bm e}}_z\right.\\
		&\left. +\curl \left[ ({\bm C} \cdot \uu)\hat{{\bm e}}_z\right]+ m ({\bm F} \cdot \uu)\cdot ({\grad T}/{T}) \times \hat{{\bm e}}_z\right].
\end{aligned}
\end{equation}
As for the existing hydrodynamic materials, GaAs quantum well satisfies the above conditions. In fact, (110)-asymmetric quantum well in GaAs possesses a crystal structure in the point group $C_{s}$ and it is believed that the circular photogalvanic effect, which is characterized by ${\bm D}$, has already been observed in the system~\cite{Moore2010}. On the other hand, ${\bm A}$ is not allowed in either of ML or BL graphene, both of which have three mirror lines in the plane. However, if we apply a uniaxial strain on the (AB-stacked) BL-graphene, we can realize finite ${\bm A}$ since the operation reduces the mirror lines from three to one. In fact, very recently, similar strategy has been executed in the system for the observation of QNHE~\cite{Ho2019}. Moreover, at least from the symmetry viewpoint, monolayer transition metal dichalcogenides (ML-TMDs) could be promising candidates for electron hydrodynamics with finite ${\bm A}$ (see also the table~\ref{table}).

\textit{Model for quantitative estimation.}---
Finally, let us consider a simple model for strained TMDs or graphene with a staggered sublattice potential~\cite{Xiao2007, Giovannetti2007} to estimate the quantitative behavior of ${\bm A}$. 
As is well-known, the Hamiltonian around the valleys $K$ and $K'$ is given by \cite{Sodemann2015}
\begin{equation}
H_\alpha = v p_x \sigma_y - \alpha v p_y \sigma_x + \alpha s p_y \1 +\Delta \sigma_z, 
\end{equation}
where $\alpha=\pm 1$ denotes the valley index at ${\bm P}_\pm=\frac{2\pi\hbar }{a}(\pm2/3,0)$. $\Delta$ is the mass gap and $s$ denotes the strain parameter. This Hamiltonian has a mirror symmetry $M_y$ and thus the hydrodynamic coefficients ${\bm A}$ must be directed in the $y$-direction.   
By diagonalizing the Hamiltonian, we can easily obtain the expression of the energy dispersion 
$\e_\alpha(\pp)=\alpha  s p_y + \mathrm{sgn}(\mu)(\Delta^2+v^2 \pp^2 )^{1/2}$ and the Berry curvature
\begin{equation}
\Omega_\alpha = \frac{\mathrm{sgn}(\mu)}{2} \frac{\alpha v^2\hbar^2\Delta}{(\Delta^2 + v^2\pp^2)^{3/2}},
\end{equation}
where $\mu>0$ ($\mu<0$) for the conduction (valence) band. Especially in the low carrier and week strain limit, we can approximate the dispersion with an isotropic parabolic form : $\e_\alpha(\pp)\simeq \Delta + (\pp+\pp_\alpha)^2/2m+ \mathcal{O}(s/v)^2$ where $m= \Delta/v^2 $ and $\pp_\alpha= (0, \alpha s \Delta/v^2) $. 
As for the Berry curvature dipole $\hat{D}$, the analytical formula at zero temperature has already been obtained in Ref.~\cite{Sodemann2015}. On the other hand, the other coefficient $\hat{C}$ can also be easily calculated in the above limit, and finally we obtain the following analytic formula: $c_x=C_{xz}=0$ and
\begin{equation}
c_y=C_{yz}=\frac{s \Delta^2}{2\pi v^2}\left[ 
\frac2\Delta - \frac{2\Delta^2 + 3(vp_F)^2}{(\Delta^2 + (v p_F)^2)^{3/2}}
\right]+ \mathcal{O}(s/v)^2
\end{equation}
where $p_F=\sqrt{2m|\mu-\Delta|}$ is the radius of the Fermi surface and, when $p_F=0$, $c_y$ becomes zero. We can estimate a typical scale of $c_y$ for ML-TMDs as $1.9\times 10^{-28}$ kg$\cdot$m/s, 
using the parameter $v\sim 4.5\times 10^5$ m/s, $\Delta\simeq 1.5$ eV, $s/v=0.1$~\cite{Sodemann2015}, and $\mu/\Delta=0.1$, which corresponds to $n\simeq 8\times 10^{13}$ cm${}^{-2}$. Using these values and $l\sim \SI{1}{\micro m}$, we can estimate the anomalous (normalized) current in the asymmetric Poiseuille flow as $J_y'\equiv J_y/E_y\sigma_D=0.02$ at $x=w/2\ (w=\SI{10}{\micro m})$, which is large enough to detect in experiments.

\textit{Conclusions.}---
In summary, we have developed a basic framework of electron hydrodynamics in noncentrosymmetric metals, which is composed of a generalized Euler equation~(\ref{Euler}) and the hydrodynamic expression of electric current~(\ref{hydro_current}). 
The obtained equation uncovers the analogy between electron fluids in crystals and chiral fluids in vacuum, and predicts various novel transport phenomena beyond linear and local responses. In particular, the generalized vortical effect, which highlights the above analogy, gives rise to unprecedented hydrodynamic transport, that is, the asymmetric Poiseuille flow and the anomalous edge current. We have confirmed that these phenomena can be found in various existing hydrodynamic materials such as WP${}_2$, BL-graphene (under strain) and GaAs quantum well, with local current probes. 

Although our formulation is based on the Fermi liquid theory, the obtained hydrodynamic theory itself might be applicable even to the strongly correlated systems beyond quasi-particle picture, since the equation is essentially no more than the continuity equation of electron momentum, which is valid irrespective of the correlation strength. It remains an intriguing problem to study what anomalous or critical behaviors the electron fluids show in noncentrosymmetric hydrodynamic materials beyond the Fermi liquid description.

\begin{acknowledgments}

We are thankful to Atsuo Shitade, Hikaru Watanabe and Akito Daido for valuable discussions. This work is supported by a Grant-in-Aid for Scientific Research on Innovative Areas: Topological Materials Science (KAKENHI Grant No. 15H05855) and also JSPS KAKENHI (Grants JP16J05078, JP18H01140 and JP19H01838). K.T. thanks JSPS for support from Research Fellowship for Young Scientists and Overseas Research Fellowship.

\end{acknowledgments}

\bibliographystyle{apsrev4-1}

\begin{thebibliography}{54}
\bibitem [{vis()}]{viscosity}%
  \BibitemOpen
  \href@noop {} {\ }\bibinfo {note} {Viscosity is defined through the momentum
  flux $\Pi_{ij}$ as $\Pi_{ij}\propto\eta_{ijkl}\partial_ku_l$ and, without
  external fields, the form agrees with the conventional one in
  Ref.\cite{Landau_Fluid} under the approximation of the isotropic parabolic
  dispersion. Although additional viscosity terms will appear through the
  anomalous velocity when we apply electric field on the fluids, they
  contribute in the order of $E\partial u\sim \mathcal{O}(E^2)$ and thus is
  negligible in the present example.}\BibitemShut {Stop}%
\bibitem [{Yip()}]{Yip}%
  \BibitemOpen
  \href@noop {} {\ }\bibinfo {note} {In this situation, it is also expected
  that the local current shows similar edge currents and thereby formulates
  magnetic flux in $x$-$z$ plane, which is in contrast with the results in
  Ref.~\cite{Yip2002} where magnetic moment induced by supercurrent is
  discussed. As for this point, further discussion may be needed in the
  future.}\BibitemShut {Stop}%
\end{thebibliography}

%

\clearpage

\renewcommand{\thesection}{S\arabic{section}}
\setcounter{section}{0}
\renewcommand{\theequation}{S\arabic{equation}}
\setcounter{equation}{0}
\renewcommand{\thefigure}{S\arabic{figure}}
\setcounter{figure}{0}

\onecolumngrid
\begin{center}
{\large {\bfseries Supplemental Materials for \\ ``Anomalous Hydrodynamic Transport in Interacting Noncentrosymmetric Metals"}}
\end{center}
\vspace{10pt}
\twocolumngrid

\section{Detailed derivation of hydrodynamic equations}

Here, we outline the derivation of the hydrodynamic equations for the TRS-preserved noncentrosymmetic metals. We start from the Boltzmann equation
\begin{equation}\label{Boltzmann}
	\pdv{f_\alpha}{t} + {\dot \rr}\pdv{f_\alpha}{\rr} + {\dot \kk} \pdv{f_\alpha}{\kk}=\mathcal{C}[f_\alpha],
\end{equation}
and the semi-classical equations~\cite{Xiao2010} 
\begin{equation}
	\hbar{\dot \kk}_\alpha =-e\E,\ \ \ 
	{\dot \rr}_\alpha=\frac1\hbar \pdv{{\e}_\alpha(\kk)}{\kk} -{\dot \kk}_\alpha\times
	{\bm \Omega}_\alpha(\kk),
\end{equation}
where $\alpha$ is a band index, $f_\alpha$ is the nonequilibrium distribution function, $-e$ is the electron charge, $\e_\alpha$ is the energy dispersion for the $\alpha$ band. The right side of Eq.~(\ref{Boltzmann}) is the scattering term, which includes both of momentum-conserving and momentum-relaxing processes.  
${\bm \Omega}_\alpha$ is the Berry curvature of the Bloch electrons, defined as ${\bm \Omega}_\alpha\equiv\grad_\kk\times {\bm A}_\alpha$ with ${\bm A}_\alpha\equiv
i\average{u_{\alpha\kk}|\grad_\kk u_{\alpha\kk}}$.
In particular, it should be noted that the Berry curvature appears only if the systems break the time-reversal or inversion symmetry. 

Following the standard approach~\cite{Landau_Kinetics, Lucas2018_review}, the continuity equations for the particle density and the momentum are obtained in the relaxation-time approximation for momentum-relaxing scattering processes as follows:
	\begin{equation}\label{particle}
		\pdv{n}{t} + \div {\bf J}_n =0
	\end{equation}
\begin{equation}\label{p_continue}
	\pdv{P_i}{t} + \pdv{\Pi_{ij}}{x_j}+ enE_i  =-\frac{P_i}{\tau_{mr}},
\end{equation}
where $n$ and ${\bm P}$  are the particle number density and the momentum of electrons respectively, and $\tau_{mr}$ is the
momentum-relaxing time. Here, ${\bm J}_n$ and $\Pi_{ij}$ are the fluxes of the particle density and the momentum, which are defined respectively as 
\begin{equation}\label{particle_flux1}
{\bf J}_n(t,\rr) = \sum_\alpha\int [d\pp] \left( \pdv{{\e}_\alpha}{\pp} +\frac{e\E}{\hbar}\times
{\bm \Omega}_\alpha \right) f_\alpha,
\end{equation}
\begin{equation}\label{momentum_flux1}
\Pi_{ij}\equiv \sum_\alpha\int [d\pp]\ p_i   \left( \pdv{{\e}_\alpha}{p_j} +\frac{e}{\hbar}\e_{jkl} E_k\Omega_{\alpha ,l} \right)f_\alpha,
\end{equation}
where we have used the notation $\int [d\pp]\equiv\int d\pp/(2\pi\hbar)^d$.

In the following, we will show the detailed derivation of Eq.~(\ref{p_continue}). Eq.~(\ref{particle}) is also derived with the same procedure. First, multiplying the Boltzmann equation (\ref{Boltzmann}) with momentum and integrating the momentum, we obtain the following equation:
\begin{equation}\label{integral}
\pdv{P_i}{t}+ \pdv{\Pi_{ij}}{x_j}-eE_j \sum_\alpha \int [d\pp] p_i \pdv{f_\alpha}{k_j} = \int  [d\pp]  p_i  \mathcal{C}[f_\alpha],
\end{equation}
where ${\bm P} \equiv \sum_\alpha\int [d\pp]  \pp f_\alpha$ and $\Pi_{ij}$ is the momentum flux, which is given in Eq. (\ref{momentum_flux1}). Performing a partial integration of the third term on the left-hand side, we can express the term with  particle number density $n\equiv \sum_\alpha \int[d\pp] f_\alpha $ as 
\begin{equation}
-eE_j \sum_\alpha \int [d\pp] p_i \pdv{f_\alpha}{k_j} = enE_i.
\end{equation}
On the other hand, we can generally decompose the scattering term into the mometum-relaxing scattering term and the momentum conserving term:
\begin{equation}
\mathcal{C}[f_\alpha] = \mathcal{C}^{mr}[f_\alpha] + \mathcal{C}^{mc}[f_\alpha].
\end{equation}
Especially in the relaxation time approximation for the momentum-relaxing scatterings, we can express the former term as
\begin{equation}
\mathcal{C}^{mr}[f_\alpha]= -\frac{f_\alpha-f_{0\alpha}}{\tau_{mr}},
\end{equation}
where we have introduced the momentum-relaxing time $\tau_{mr}$ and the local Fermi distribution function~: $f_{0\alpha}(\xx,\pp,t)\equiv [1+e^{-\beta(\e_\alpha(\pp)-\mu )}]^{-1}$. Furthermore, it is well-known that the integration of the product of a scattering term and a conserved quantity in the scattering process is always zero ~\cite{Landau_Kinetics, Lucas2018_review} and thus, especially for the momentum, $\int  [d\pp]  p_i  \mathcal{C}^{mc}[f_\alpha]=0$. Consequently, we can evaluate the right-hand side of Eq.~(\ref{integral}) as
\begin{equation}
\int  [d\pp]  p_i  \mathcal{C}[f_\alpha] = -\frac{P_i}{\tau_{mr}},
\end{equation}
and finally we obtain the continuity equation for momentum (\ref{p_continue}). 

In the hydrodynamic regime, the most essential assumption is that we can describe the distribution functions with the perturbation theory from the following {\it local equilibrium function}~\cite{Landau_Kinetics, Lucas2018_review}
\begin{equation}\label{local_equilibrium}
f_{\uu\alpha}(\xx,\pp,t)\equiv \frac{1}{1+e^{-\beta(\e_\alpha(\pp)-{\bm u}\cdot\pp-\mu )}}.
\end{equation}
Here, $\beta$, $\mu$ and ${\bm u}$ are the spatiotemporal functions specifying the hydrodynamic variables: (inverse) temperature, chemical potential and velocity. 
In the following discussion, we further assume that each band structure can be approximated by an isotropic parabolic dispersion with the same effective mass $m$ around some valleys: $\e_\alpha(\pp) \simeq \pp^2/2m$, where $\pp$ is defined as a deviation from the valley. For example, when considering graphene with inversion breaking or ML-TMDs such as MoS${}_2$, the centers of valleys correspond to $K$ and $K'$ point in the Brillouin zone and the above condition is satisfied. On the other hand, hydrodynamic materials with linear dispersion, such as WP${}_{2}$, do not satisfy this condition for the dispersion. In particular, WP${}_2$ is considered as a Weyl semimetal and thus could show a peculiar hydrodynamic flow due to the chiral anomaly. However, we believe that the anomalous hydrodynamic flows predicted by our theory will be realized even in these materials, since our analysis in the main text partially relies on symmetry consideration of the hydrodynamic materials. Of course, to clarify the role of the linear dispersion or the chiral anomaly at the microscopic level, further discussions beyond our theory will be needed.

Under the above assumptions, for example, we can estimate the momentum density in the zeroth-order approximation for the distribution function as
\begin{equation}\label{S9}
\begin{split}
{\bm P} &\equiv \sum_\alpha\int [d\pp]  \pp f_\alpha \simeq  \sum_\alpha\int [d\pp]  \pp f_{\uu\alpha}\\
& = \sum_\alpha\int [d\pp]  (\pp+m\uu) f_{0\alpha}(\pp) +\mathcal{O}(u^3)\\
& = mn{\bm u}  +\mathcal{O}(u^3),
\end{split}
\end{equation}
where we have performed a variable transformation $\pp\to\pp+m\uu$ in the second line and used the fact that $f_{\uu\alpha}(\pp+m\uu)=f_{0\alpha}(\pp)+\mathcal{O}(u^3)$ due to the parabolic dispersion. 
Consequently, the first of the particle flux (\ref{particle_flux1}) can be described in terms of the velocity field as 
\begin{equation}\label{eq8}
\begin{split}
 &\sum_\alpha\int [d\pp]  \pdv{{\e}_\alpha}{\pp}f_\alpha(\pp) \\ 
&\simeq \frac{1}{m}\sum_\alpha \int[d\pp] \pp  f_{\uu\alpha}(\pp) 
=\frac{\bm P}{m}= n{\bm u}  +\mathcal{O}(u^3).
\end{split}
\end{equation}
On the other hand, the second term of the particle flux are calculated as follows:
\begin{equation}\label{S11}
\begin{split}
&\sum_\alpha\int [d\pp] \frac{e}{\hbar}\e_{ijk}E_j {\Omega}_{\alpha,k}  f_\alpha \\
&\simeq \sum_\alpha\int [d\pp] \frac{e}{\hbar}\e_{ijk}E_j {\Omega}_{\alpha,k}  f_{\uu\alpha} \\
&= \sum_\alpha\int [d\pp] \frac{e}{\hbar}\e_{ijk}E_j {\Omega}_{\alpha,k}(\pp + m\uu) f_{0\alpha}+ \mathcal{O}(E^3)\\
&= \frac{em}{\hbar} \sum_\alpha   \e_{ijk}E_j u_l  \int [d\pp]  
\pdv{ \Omega_{\alpha,k}}{p_l}  f_{0\alpha}  +\mathcal{O}(E^3), 
\end{split}
\end{equation}
where, in the third line, we have used the transformation $\pp\to\pp+m\uu$ and the fact that the dispersion can be approximated as a parabolic one. Furthermore, in the final line, we have expanded the Berry curvature in powers of $m\uu$ and used the fact that the Berry curvature is an odd function of momentum around the $\Gamma$ point due to the TRS, which means that the sum of the integral of ${\bm \Omega}_\alpha$ at each valley always becomes zero: $\sum_\alpha \int [d\pp]{\bm \Omega}_\alpha f_\alpha=0$. Moreover, in the both lines, we assume that the velocity field $\uu$ is driven by an electric field $\E$ and drop the terms with more than third order in $\uu$ and $\E$. Finally, we end up with the hydrodynamic expression of the particle flux, which are correct up to the second order in $\uu$ and $\E$, as follows:
\begin{equation}
	{\bm J}_n = n{\bm u} +\frac{em}{\hbar} \E\times {}^t\!\hat{D} {\bm u},
\end{equation}
where the symbol $t$ denotes the transpose of a matrix and $D_{il}$ is a geometrical coefficient, so-called Berry curvature dipole, defined as 
\begin{equation}
D_{il} = \sum_\alpha D_{il}^\alpha ,\ \  D_{il}^\alpha \equiv -\int [d\pp] \Omega_{\alpha,l} \pdv{f_{0\alpha}}{p_i},
\end{equation}

As for the momentum flux (\ref{momentum_flux1}), by performing similar procedures, we can obtain the hydrodynamic expression. First, we can decompose the first term of Eq. (\ref{momentum_flux1}) into two terms as 
\begin{equation}\label{S14}
\begin{split}
\sum_\alpha\int [d\pp]\ &p_i  \pdv{{\e}_\alpha}{p_j}f_\alpha=\frac1m \sum_\alpha\int [d\pp]\ p_i  p_j f_\alpha\\ 
&= \frac1m \sum_\alpha\left[\frac{1}{n_\alpha}\average{p_i}_\alpha\average{p_j}_\alpha+ \average{\delta p_i \delta p_j}_\alpha\right]\\
\end{split}
\end{equation}
where $\average{\cdots}_\alpha\equiv \int [d\pp] (\cdots)f_\alpha$, $n_\alpha\equiv \int[d\pp]f_\alpha$ and $\delta \pp \equiv\pp-\average{\pp}_\alpha/n_\alpha$. From the calculation in Eq.~(\ref{S9}), we obtain $\average{\pp}_\alpha=mn_\alpha \uu$ and thus, the first term of Eq.~(\ref{S14}) can be described in terms of the velocity field as
\begin{equation}
 \frac1m \sum_\alpha \frac{1}{n_\alpha}\average{p_i}_\alpha\average{p_j}_\alpha 
= mnu_iu_j
\end{equation}
Next, we estimate the the second term of Eq.~(\ref{S14}) can be estimated as follows. As can be easily seen, the off-diagonal term of $\average{\delta p_i \delta p_j}_\alpha$ is always zero due to the isotropy of the band around each valley: $\average{\delta p_i \delta p_j}_\alpha=0 $ ($i\neq j$). On the other hand, the diagonal terms of $\average{\delta p_i \delta p_j}_\alpha$ are calculated as follows:
\begin{equation}
\begin{split}
\frac1m\sum_\alpha\average{\delta p_i \delta p_i}_\alpha &=\frac1m\sum_\alpha\int[d\pp] (p_i - m_\alpha u_i)^2 f_{\uu\alpha}(t,\rr,\pp)\\
&=\frac1m\sum_\alpha\int[d\pp] p_i^2 f_{0\alpha}(t,\rr,\pp) \\
&=\frac1m\sum_\alpha\frac1d\int [d\pp]\pp^2 f_{0\alpha}(t,\rr,\pp) \\
&=\frac1m\sum_\alpha\frac{2m}{d}\int [d\pp]\e_\alpha(\pp) f_{0\alpha}(t,\rr,\pp)\\
&=\frac{2}{d}\e(\rr,t) = P
\end{split}
\end{equation}
where, in the first line, we have used the relation $\delta \pp \equiv\pp-\average{\pp}_\alpha/n_\alpha=\pp-m\uu$ and, in the second line, we have performed the transformation $\pp\to\pp+m\uu$. Furthermore, in the final equality, we have applied the familiar formula between the energy density $\e$ and the pressure $P$ for the Fermi gas system: $P=2\e/d$~\cite{Fetter}.

Next, we estimate the second term of Eq.~(\ref{momentum_flux1}). Performing the usual transformation $\pp\to\pp+m\uu$, this term can be transformed as
\begin{equation}
\begin{split}
&\frac{e}{\hbar}\e_{jkl} E_k\sum_\alpha\int [d\pp]\Omega_{\alpha ,l}(\pp) f_\alpha(\pp)\\
&=\frac{e}{\hbar}\epsilon_{jkl}E_k\sum_\alpha\int [d\pp] (p_i+m_\alpha u_i)  \Omega_{\alpha,l}(\pp+m_\alpha \uu)f_{0\alpha}(\pp).
\end{split}
\end{equation}
As with the procedure in Eq.~(\ref{S11}), by expanding this term in powers of $\uu$, we can approximate it within the second order in $\uu$ and $\E$ as follows: 
\begin{equation}\label{S18}
\begin{split}
&=\frac{e}{\hbar}\epsilon_{jkl}E_k  \sum_\alpha\int [d\pp] (p_i+m_\alpha u_i)\\
&\ \ \ \ \ \ \ \times\left(\Omega_{\alpha,l}+m_\alpha u_n  \pdv{\Omega_{\alpha,l}}{p_n}\right)  f_{0\alpha}(\pp)+\mathcal{O}(E^3)\\
&= \frac{e}{\hbar}\epsilon_{jkl}E_k  \sum_\alpha\int [d\pp] p_i  \Omega_{\alpha,l} f_{0\alpha}(\pp)+\mathcal{O}(E^3)\\
&= \frac{e}{\hbar}\epsilon_{jkl} C_{il} E_k+\mathcal{O}(E^3),
\end{split}
\end{equation}
where $C_{il}$ is the second geometrical coefficient, defined as
\begin{equation}
C_{il} = \sum_\alpha C_{il}^\alpha ,\ \  C_{il}^\alpha \equiv \int [d\pp]\ p_i  \Omega_{\alpha,l} f_{0\alpha}
\end{equation}
In the second line, we have used the fact that the Berry curvature is an odd function of momentum around the $\Gamma$ point due to the TRS and, in addition, dropped the term proportional to $E_k u_iu_n$ since it is in the order of $E^3$.

Finally, substituting Eqs.~(\ref{S14})-(\ref{S18}) into Eq.~(\ref{momentum_flux1}), we end up with the hydrodynamic expression of the momentum flux, which are correct up to the second order in $\uu$ and $\E$:
\begin{equation}\label{momentum_flux}
	 \Pi_{ij}=
		mn u_iu_j+
		P\delta_{ij}+
		 \frac{e}{\hbar}\epsilon_{jkl}C_{il} E_k,
\end{equation}
We note that, at least in our assumption for band dispersions, it is shown that the trace component of $\hat{C}$ always vanishes because the Berry curvature is divergence free. This can be easily seen as follows:
\begin{equation}
\begin{split}
C\equiv \mathrm{Tr}[\hat{C}]&=\int[d\pp] p_i \Omega_{\alpha,i} f_{0\alpha}\\
&\propto\int[d\pp] \pdv{\e}{p_i} \Omega_{\alpha,i} f_{0\alpha}\\
&=\int d\e f_{0\alpha}(\e) \oint_{\e_{\alpha}(\kk)=\e} (d{\bm S}\cdot \Omega_{\alpha}(\kk)) \\
&=0.
\end{split}
\end{equation}
Here, at the third line, we have performed the variable transformation $d\kk=dSdk_{\perp}=\frac{dS}{|\grad_\kk \e|}d\e$, where $dS$ is the area element on the isoenergy surface with $\e_\alpha(\kk)=\e$. Furthermore, at the final equation, we have used the fact that the Berry curvature is divergence free. 

Substituting the above expression into the continuity equations~(\ref{p_continue}) and (\ref{particle}), we obtain the desired hydrodynamic equations for noncentrosymmetric metals, in the approximation up to the second order in $\uu$ and $\E$:
\begin{equation}
\pdv{n}{t}+\div(n{\bm u}) +\frac{em}{\hbar}\div(\E\times {}^t\!\hat{D} {\bm u})=0
\end{equation}
\begin{equation}\label{Euler1}
\begin{split}
		&\pdv{u_i}{t}+({\bm u}\cdot \grad)u_i +\frac{1}{\rho}\pdv{p}{x_i} + \\
		& \frac{e}{mn\hbar}\epsilon_{jkl}  \left[C_{il}\pdv{E_k}{x_j}
				+\pdv{C_{il}}{x_j}E_k \right]+ \frac{e}{m}E_i  =-\frac{u_i}{\tau_{mr}},
\end{split}
\end{equation}
Furthermore, the derivative of the coefficient $C_{il}^\alpha$ can be estimated explicitly as 
\begin{equation}
\begin{split}\label{gradC}
\pdv{C_{il}^\alpha}{x_j} &= \int [d\pp] p_i  \Omega_{\alpha,l} \pdv{}{x_j} f_{0}(t,\rr,\pp)\\
&= \int [d\pp] p_i  \Omega_{\alpha,l} \left[-\frac{1}{T^2}\pdv{T}{x_j}\pdv{f_{0\alpha}}{\beta}+\pdv{\mu}{x_j}\pdv{f_{0\alpha}}{\mu}\right]\\
&=\frac{mF_{il}^\alpha}{T}\pdv{T}{x_j} + m D_{il}^\alpha\pdv{\mu}{x_j}
\end{split}
\end{equation}
where we have defined the third geometrical coefficients as 
\begin{equation}
F_{il}^\alpha \equiv -\int [d\pp] p_i \e_\alpha  \Omega_{\alpha,l} \pdv{f_{0\alpha}}{\e}
\end{equation}
Substituting this formula into Eq.~(\ref{Euler1}), we reach the generalized Euler equation Eq.~(2) in the main text.

Next, we move on to the derivation of the hydrodynamic expression for the transport current.  
Generally, the current is known to be described with orbital magnetization ${\bm M}$ in the following form~\cite{Cooper1997, Xiao2006}:
\begin{equation}
{\bm J}={\bm J}_n + \curl \left(\sum_\alpha \int[d\pp]{\bm m}_\alpha f_\alpha\right) -\curl {\bm M}.
\end{equation}
Especially in the equilibrium state, the orbital magnetization ${\bm M}$ is expressed as
\begin{equation}\label{orbital magnetization}
\begin{split}
{\bm M} = &\sum_\alpha \int[d\pp]\ {\bm m}_\alpha f_\alpha
+ \sum_\alpha \frac1\beta \\
&\times  \int [d\pp]\ \frac e\hbar {\bm \Omega}_\alpha \cdot \log (1+e^{-\beta(\e_\alpha-\mu)}).
\end{split}
\end{equation}
In this study, to estimate the orbital magnetization in the local-equilibrium state, we phenomenologically assume that the microscopic expression is obtained by replacing the exponent $\e_\alpha$ in Eq.~(\ref{orbital magnetization}) with $\e_\alpha-{\bm u}\cdot\pp$. Although this prescription is kind of phenomenological, the resulting formula~(6) in the main text seems to support the validity, because it guarantees the equivalence between the external electric field $\E$ and statistical force $\grad \mu$ for the transport current. 

Under this assumption, we can express the transport current as 
\begin{equation}
\begin{split}
		{\bm j}&=-e \sum_\alpha\int \dot{\rr}_\alpha f_\alpha(t,\rr,\pp) \frac{d\pp}{(2\pi\hbar)^d}\\
		&-\sum_\alpha\curl \frac1\beta \int [d\pp]\frac e\hbar {\bm \Omega}_\alpha(\pp)\cdot \log (1+e^{-\beta(\e_\alpha-{\bm u}\cdot\pp-\mu)}).
\end{split}
\end{equation}
For the integration in the second term, we can estimate it up to the second order in $\uu$:
\begin{equation}
	\begin{split}
		&\frac1\beta\int [d\pp]\frac e\hbar {\bm \Omega}_\alpha(\pp)\cdot \log (1+e^{-\beta(\e_\alpha-{\bm u}\cdot\pp-\mu)})\\
		&=\int [d\pp]\frac e\hbar {\bm \Omega}_\alpha(\pp)(\uu\cdot \pp)f_0(t,\rr,\pp)+\mathcal{O}(u^3)\\
&\simeq \frac e\hbar {}^t\!\hat{C}^\alpha(\rr) {\bm u}(\rr),
	\end{split}
\end{equation}
where we have used the fact that the Berry curvature is an odd function of the momentum around the $\Gamma$ point due to the TRS. Finally, using the formula~(\ref{gradC}), we reach the hydrodynamic expression for the transport current (6) in the main text.  
Here, we note that the above expression of the current does not include the full intrinsic contributions with the first-order of spacial gradient. 
This is because, in the derivation of Eq.~(6) in the main text, we drop the contribution of the first-order deviation of the non-equilibrium distribution function from Eq.~(\ref{local_equilibrium}). Consequently, to recover the dropped contributions such as the Fourier's law for thermal gradient, we must proceed to the calculation of the next order perturbation, which is roughly proportional to the electron-electron scattering time $\tau_{ee}$ and thus negligible in the strong correlation limit ($\tau_{ee} \to 0$).

\providecommand{\noopsort}[1]{}\providecommand{\singleletter}[1]{#1}%

\end{document}